\documentclass[twocolumn,english,aps,pra,floatfix,tightenlines,superscriptaddress]{revtex4-1}
\usepackage[T1]{fontenc}
\usepackage[latin9]{inputenc}
\usepackage{babel}
\usepackage{amsmath}
\usepackage{amssymb}
\usepackage{graphicx}
\usepackage{esint}
\usepackage{subfigure}

\usepackage[pdfstartview=FitH]{hyperref}
\hypersetup{
    colorlinks=true,       
    linkcolor=cyan,          
    citecolor=magenta,        
    filecolor=magenta,      
    urlcolor=cyan,           
    runcolor=cyan
}
\usepackage[pdftex]{color}

\newcommand{\braket}[2]{\langle #1 | #2\rangle}
\newcommand{\ket}[1]{ | #1\rangle}
\newcommand{\bra}[1]{\langle #1 | }

\newcommand{\tr}{\mathrm{Tr}}

\newcommand{\1}{\openone}

\newcommand{\beq}{\begin{equation}}
\newcommand{\eeq}{\end{equation}}
\newcommand{\bes}{\begin{subequations}}
\newcommand{\ees}{\end{subequations}}
\newcommand{\bea}{\begin{eqnarray}}
\newcommand{\eea}{\end{eqnarray}}
\newcommand{\ignore}[1]{}

\DeclareMathOperator{\mydash}{\!-\!}

\makeatother

\begin{document}

\title{Adiabaticity in open quantum systems}

\author{Lorenzo Campos Venuti}
\affiliation{Department of Physics \& Astronomy, University of Southern California, Los Angeles, CA 90089, USA}
\affiliation{Center for Quantum Information Science \& Technology, University of Southern California, Los Angeles, CA 90089, USA}

\author{Tameem Albash}
\affiliation{Department of Physics \& Astronomy, University of Southern California, Los Angeles, CA 90089, USA}
\affiliation{Center for Quantum Information Science \& Technology, University of Southern California, Los Angeles, CA 90089, USA}
\affiliation{Information Sciences Institute, University of Southern California, Marina del Rey, California 90292, USA}

\author{Daniel A. Lidar}
\affiliation{Department of Physics \& Astronomy, University of Southern California, Los Angeles, CA 90089, USA}
\affiliation{Center for Quantum Information Science \& Technology, University of Southern California, Los Angeles, CA 90089, USA}
\affiliation{Department of Electrical Engineering, University of Southern California, Los Angeles, CA 90089, USA}
\affiliation{Department of Chemistry, University of Southern California, Los Angeles, CA 90089, USA}

\author{Paolo Zanardi}
\affiliation{Department of Physics \& Astronomy, University of Southern California, Los Angeles, CA 90089, USA}
\affiliation{Center for Quantum Information Science \& Technology, University of Southern California, Los Angeles, CA 90089, USA}

\begin{abstract}
We provide a rigorous generalization of the quantum adiabatic theorem for open systems described by a Markovian master equation with time-dependent Liouvillian $\mathcal{L}(t)$. We focus on the finite system case relevant for adiabatic quantum computing and quantum annealing. Adiabaticity is defined in terms of closeness to the instantaneous steady state. While the general result is conceptually similar to the closed system case, there are important differences. Namely, a system initialized in the zero-eigenvalue eigenspace of $\mathcal{L}(t)$ will remain in this eigenspace with a deviation that is inversely proportional to the total evolution time $T$. In the case of a finite number of level crossings the scaling becomes $T^{-\eta}$ with an exponent $\eta$ that we relate to the rate of the gap closing. 
For master equations that describe relaxation to thermal equilibrium, we show that the evolution time $T$ should be long compared to the corresponding minimum inverse gap squared of $\mathcal{L}(t)$. Our results are illustrated with several examples. 
\end{abstract}


\maketitle

\section{Introduction}

The origins of the celebrated quantum adiabatic theorem (QAT) date back
to Einstein's ``Adiabatenhypothese'': ``If a system be affected
in a reversible adiabatic way, allowed motions are transformed into
allowed motions'' \cite{Einstein:adiabatic}. Ehrenfest was the first
to appreciate the importance of adiabatic invariance, guessing---before
the advent of a complete quantum theory--- that quantum laws would
only allow motions which are invariant under adiabatic perturbations
\cite{Ehrenfest:adiabatic}. The more familiar, modern version of the
QAT was put forth by Born and Fock already in 1928 for the case of
discrete spectra~\cite{born_beweis_1928}. Since then a series of
increasingly sophisticated techniques have been developed in order
to generalize the QAT to include degeneracy, unbounded models, continuous
spectra, and exponential error estimates~\cite{kato_adiabatic_1950,Garrido:62,avron:87,nenciu_linear_1993,avron_adiabatic_1999,Hagedorn:2002kx,Jansen:07,lidar:102106}.

This long history of adiabatic theorems is almost exclusively concerned with closed systems undergoing unitary evolution. Previous approaches to formulating an adiabatic condition for open quantum systems \cite{Breuer:book} have focused on a Jordan block decomposition of the dissipative generator \cite{PhysRevA.71.012331}, the weak coupling limit \cite{PhysRevA.72.022328,ABLZ:12-SI}, zero temperature \cite{Pekola:2010oj}, or on a noiseless subsystem decomposition \cite{oreshkov_adiabatic_2010,comment-SL-OC}. Here we prove that in analogy to the closed system case, where the system follows the instantaneous (pure) eigenstates of the Hamiltonian in the adiabatic limit of arbitrarily large total evolution time $T$, the open system follows the instantaneous steady state (ISS) of the Liouvillian. In doing so, we extend the seminal closed system result by Kato \cite{kato_adiabatic_1950} to infinite order in $1/T$. 

Rigorous extensions of the adiabatic theorem for generators of contractive
semigroups, similar to ours, have also appeared in the mathematical literature \cite{springerlink:10.1007/BF01011696,joye_general_2007,salem_quasi-static_2007,Avron:2012tv}. Our focus is on estimating the adiabatic error in terms of the physical parameters of the theory, thus making the result more suitable for applications. We demonstrate
that in the case of thermal baths satisfying the Kubo-Martin-Schwinger (KMS) condition \cite{KMS}, where
the ISS is the instantaneous Gibbs state, a sufficient criterion
for adiabaticity is $T\gg\Delta_{\min}^{-2}$, where $\Delta_{\min}$
is the smallest \emph{Liouvillian} gap in absolute value. Our QAT
also allows for a finite number of level crossings in the Liouvillian
spectrum, for which we demonstrate that the error scales asymptotically
as $T^{-\eta}$ with a known exponent $\eta\in(0,1)$ related to the rate
at which the gap closes. 
This setting is directly relevant to recent theoretical
and experimental work on quantum annealing \cite{ABLZ:12-SI,q-sig,DWave-16q,Boixo:2014yu},
where the Liouvillian gap may close at the end of the evolution, 
and we verify this prediction using numerical
simulations. 

\section{Instantaneous steady states}

We assume that the evolution
of a $d$-dimensional ($d<\infty$) quantum system with state $\rho(t)$ can be described by a linear, time-local
master equation $d\rho/dt=\mathcal{L}_{T}(t)\rho$, where
$T$ is the total evolution time. We also
assume that $\mathcal{L}_{T}(sT)=\mathcal{L}(s)$ where $s=t/T\in[0,1]$
is a rescaled, dimensionless time coordinate, and $\mathcal{L}(s)$
is $T$-independent. Setting $\rho_{T}(s)=\mathcal{E}_{T}(s,0)\rho_{T}(0)$,
the evolution operator $\mathcal{E}_{T}(s,s_{0})$ satisfies 
\begin{equation}
\mathcal{E}_{T}'(s,s_{0})=T\mathcal{L}(s)\mathcal{E}_{T}(s,s_{0})\ ,\label{eq:ME}
\end{equation}
with $\mathcal{E}(s,s)=\1$ {(}we drop the subscript $T$ from now
on; we also write $\mathcal{E}(s)$ for $\mathcal{E}(s,0)$ for simplicity{)}, and where the prime  denotes $\partial_s$.
We are interested in the solutions of Eq.~(\ref{eq:ME}) for large $T$. We further assume that the Liouvillian $\mathcal{L}(s)$ can be written
in Lindblad form for all $s$, i.e., 
$\mathcal{L}(s)\bullet=-i[H(s),\bullet]+\sum_{l} \left [ L_{l}(s)\bullet L_{l}^{\dagger}(s)-\frac{1}{2}\{L_{l}^{\dagger}(s)L_{l}(s),\bullet\} \right ]$,
where $H(s)$ is the system Hamiltonian and $\{L_{l}(s)\}$ are the
Lindblad operators.  Equation~\eqref{eq:ME} then describes a Markovian
master equation with a time-dependent Lindblad generator, and the
corresponding evolution operator $\mathcal{E}(s_{2},s_{1})$ is a
completely positive trace preserving (CPTP) map for any $s_{2}\ge s_{1}$
\cite{Wonderen:2000yi,Breuer:2004wq,ABLZ:12-SI}.
We formulate our results in terms of  Lindblad operators and CPTP maps, but in fact all our results are valid in the more general case where $\mathcal{L}(s)$ generates a contractive semigroup (i.e., $\parallel e^{t\mathcal{L}(s)} \parallel \le 1$ $\forall s$ and $t>0$).
A special role is played by
the ISSs, i.e., the states in the kernel of $\mathcal{L}(s)$. In the time-independent
case [$\mathcal{L}(s)=\mathcal{L}$, $\forall s$], it follows from
the CPTP property that any initial state evolves to $\mathrm{Ker}\mathcal{L}$
in the long time limit \cite{Alicki:87}. Let us denote by $P(s)$
the (instantaneous) spectral projection of $\mathcal{L}(s)$
with eigenvalue zero. The Lindblad form guarantees that zero is a semi-simple, possibly
degenerate, eigenvalue of $\mathcal{L}(s)$ (see Appendix \ref{app:0}
for a proof), and so there are no nilpotent terms in the zero sector, i.e., $\mathcal{L}(s)P(s)=P(s)\mathcal{L}(s)=0$. 

We are now ready to informally state the QAT for open
systems: \emph{If a system is initialized at $s=0$ in $\mathrm{Ker}\mathcal{L}(0)$,
the final state at $s=1$ will be close to $\mathrm{Ker}\mathcal{L}(1)$,
provided the Lindbladian changes sufficiently slowly}. 

In principle one could formulate an open-system QAT considering other (non-zero) eigenvalues
of $\mathcal{L}(s)$. However their corresponding
invariant subspaces contain no physical states, so that the physical
interest in such a generalization is questionable %
\footnote{To see this, let
$P_{0}$ be the projector onto the zero eigenvalue. Since it can be
realized as the infinite time limit of a CPTP map, $P_{0}$ is itself
a CPTP map, and in particular it is trace preserving. Let $P_{j}$
be the projector onto another invariant subspace, with $j\neq0$.
Assume there is a state in its range, i.e., $\exists\, x\,|$ $\rho=P_{j}x$
is a state. But $P_{0}P_{j}=0$ so $\tr\left(\rho\right)=\tr\left(P_{0}\rho\right)=0$
(where the first equality holds since $P_{0}$ is a CPTP map), a contradiction.%
}. 
We proceed to rigorously establish the QAT and identify the timescales it entails.

\section{Gapped case}

We start by assuming that the zero eigenvalue
is separated by a finite gap $\Delta_{\mathrm{min}}$ from the rest of the spectrum $\sigma(\mathcal{L}(s))$
for all $s\in[0,1]$, i.e.,~$\mathrm{dist}\left[\sigma(\mathcal{L}(s))\backslash\{0\},0\right] = \Delta (s) \ge \Delta_{\min} >0$,
a condition we relax later. The ideal adiabatic evolution is represented by an operator $V(s)$ that satisfies the intertwining
property: $V(s)P(0)=P(s)V(s)$. It is well known \cite{kato_adiabatic_1950}
(see Appendix.~\ref{app:intertwiner})
that a possible choice for $V(s)$ is given
by the solution of the differential equation $V'(s)=[P'(s),P(s)]V(s)$
with $V(0)=\1$. We are interested in quantifying the deviation of the actual evolution,
governed by the CPTP map $\mathcal{E}(s)$, from the ideal adiabatic evolution. However, 
$V(s)$ is not CPTP  in general. Instead, we can prove that $W(s):=V(s)P(0)$ is a CPTP map, since
it can be written as a product of projectors (CPTP maps): $W(s)=\lim_{N\to\infty}P(s)\cdots P(2s/N)P(s/N)P(0)$
(see Appendix \ref{app:positivity}) or Proposition 3 in \cite{Avron:2012tv}). 
Therefore, to state the QAT we wish to bound the deviation from the ideal
adiabatic evolution projected to $\mathrm{Ker}\mathcal{L}$, $\|\mathcal{E}(s)P(0)-V(s)P(0)\|$, in the large $T$ limit \footnote{Unless explicitly noted otherwise, from
hereon the norm is the induced trace norm, i.e., $\protect\| A\protect\|=\sup_{x\neq0}\protect\|A(x)\protect\|_{1}/\protect\|x\protect\|_{1}$,
where $\protect\| x\protect\|_{1}$ denotes the trace norm, i.e., the sum of the singular
values \cite{Bhatia:book}. For a discussion of the properties of the induced trace norm see,
e.g., Ref.~\cite{PhysRevA.78.012308}, where it is denoted $\protect\| \ \protect\|_{\infty,1}$,
or Ref.~\cite{Kretschmann:08}}.

To proceed we introduce the reduced resolvent $S(s)=\lim_{z\to0}Q(s)(\mathcal{L}(s)-z)^{-1}Q(s)$, where $Q(s)=\1-P(s)$ \cite{QAT-comment-smoothL}. We further assume that $\mathcal{L}$ is $m$ times differentiable, and let $X_{n+1}(s)  =S(s) X_{n}'(s)$, with $X_{1}(s)=S(s)$.
Under the additional simplifying assumption that the ISS is unique we then prove the following in Appendix \ref{sec:arb_order}
using integration by parts:
\begin{align}
\label{eq:adia_final}
&[\mathcal{E}(s)-V(s)]P(0) =\sum_{n=1}^{m}\frac{\Omega_{n}}{T^{n}}  \\
& \qquad \qquad \qquad\qquad -\frac{1}{T^{m}}\int_{0}^{s}\!d\sigma\,\mathcal{E}(s,\sigma)X_{m}'(\sigma)W'(\sigma) \ ,\notag \\
&\Omega_{n} \!=\!\left.\mathcal{E}(s,\sigma)X_{n}(\sigma)W'(\sigma)\right|_{0}^{s} -\int_{0}^{s}\!d\sigma \mathcal{E}(s,\sigma)X_{n}(\sigma)W''(\sigma) \ . \notag 
\end{align}
The general result, valid also for degenerate kernels, is given in Appendix \ref{sec:degenerate_kernel}. 

It turns out that Eq.~\eqref{eq:adia_final} is valid in the $m=1$ case even without requiring that the ISS be unique, and as we show in Appendix \ref{sec:arb_order}
we can bound the deviation from the ideal adiabatic evolution in general as
\begin{equation}
\| \left[\mathcal{E}(s)-V(s)\right]P(0)\| \le C/T\ .
\label{eq:adia_bound}
\end{equation}
A similar result has been derived in \cite{Avron:2012tv} where, however, the constant $C$ is left undetermined. We show that the constant  $C$, independent of $T$, can be taken to be
\begin{multline}
C=\| S(s)\| \| P'(s)\| +\| S(0)\| \| P'(0)\| \\
+\sup_{\sigma\in[0,s]}\| [S'P'+SP''](\sigma)\| \ .
\label{eq:const_bound}
\end{multline}
Below we discuss how $C$ relates to the physical parameters of the model associated with $\mathcal{L}$. \emph{Inequality} \eqref{eq:adia_bound} \emph{states the QAT for open systems and implies the QAT in the standard form for states}. 

To see the latter let us initialize the system in a state
$\tilde{\rho}(0)$ in $\mathrm{Ker}\mathcal{L}(0)$, i.e., $\tilde{\rho}(0)=P(0)\tilde{\rho}(0)$.
Then $\tilde{\rho}(s):=V(s)\tilde{\rho}(0)$ is an instantaneous steady
state at time $s$, since $\mathcal{L}(s)\tilde{\rho}(s)=\mathcal{L}(s)V(s)P(0)\tilde{\rho}(0)=\mathcal{L}(s)P(s)V(s)\tilde{\rho}(0)=0$,
i.e., $\tilde{\rho}(s)\in\mathrm{Ker}\mathcal{L}(s)$. Under the actual
evolution the state is mapped to $\rho(s)=\mathcal{E}(s)\tilde{\rho}(0)$
and one has $\| \rho(s)-\tilde{\rho}(s)\|_1 \le\| \left[\mathcal{E}(s)-V(s)\right]P(0)\| \| \tilde{\rho}(0)\|_1 \le C/T$.
Namely, if $T\ge C/\epsilon$ ($\epsilon>0$) then the system is guaranteed to find itself $\epsilon$-close in norm to the instantaneous
steady state, at the end of the evolution.%

\section{The Closed System Limit}
It is useful to comment on how our result relates to adiabatic theorems for closed systems, described by a system Hamiltonian $H(s)$ with eigenvalues $E_n$.

First, if one is interested in initial states belonging to the $-i E_0$ level, one may simply set ${\cal L}(s) = -i [H(s)-E_0]$, as our formalism encompasses (with minor modifications) the case where ${\cal L}$ is anti-Hermitian. In this way one recovers the standard adiabatic theorem for closed system. The relevant gap is given by the eigenvalue closest to $E_0$ in modulus, i.e., $|E_1 -E_0|$. The bound we obtained for the constant $C$ in this case is similar to that given in Ref.~\cite{Jansen:07}, at least for what concerns the dependence on the gap.

Another possibility is to write ${\cal L}(s)= {\cal K}(s)= -i[H(s),\bullet]$. The eigenvalues of ${\cal K}$ are $\{-i (E_n -E_m)\}$. There is a $\ge d$-fold degenerate zero eigenvalue arising from $E_n=E_m$ constituting ${\rm Ker} {\cal K}$. The relevant energy scale  is determined by the next eigenvalue which is closest to zero in modulus. This is given by the smallest difference $| E_{n} -E_{m} |$ with $E_n\neq E_m$ (non-zero since we assume $d<\infty$ and hence discrete spectra). This is consistent with the previous result because in ${\rm Ker} {\cal K}$ one has the freedom to pick any state $| n\rangle \langle n |$ leading to a gap $\min_{m} |E_{n} -E_{m}| $. In this manner one obtains an adiabatic theorem for closed systems in the Liouvillian (superoperator) formalism. We discuss the closed system limit further in Appendix~\ref{app:closed}.

\section{Thermal bath}

It turns out that the open system version of the QAT can have additional
structure that is absent in the closed system case. To demonstrate
this we consider the important class of Lindbladians generated by
the interaction of a system with a thermal bath, for which we can
make the bounds above more specific. As a result of the KMS condition
such Lindbladians satisfy the quantum detailed balance condition \cite{alicki_detailed_1976,Kossakowski:1977dk}.
This fact has important consequences, namely, (i) the Gibbs
state is an ISS, i.e., $\mathcal{L}(s)\rho_{G}(s)=0$
with $\rho_{G}(s)\equiv\exp\left[-\beta H(s)\right]/Z$,
where 
$Z=\tr\exp\left[-\beta H(s)\right]$ is the partition function; (ii) the generator $\mathcal{L}(s)$ is normal.

Let us now show how we can relate $C$ to standard quantities such as the gap and $H'(s)$ using the assumption of a thermal bath. Assume for simplicity that $\rho_{G}(s)$ is the unique state
in $\mathrm{Ker}\mathcal{L}(s)$. For such thermal baths the
projector onto the ISS manifold is $P(s) = |\rho_G (s) \rangle \langle \1 |$ 
Then 
$P^{(n)}(s)=|\rho_G^{(n)} (s) \rangle \langle \1 |$, so that
$\|P^{(n)}(s)\|=\|\rho_{G}^{(n)}(s)\|_{1}$. Thus, if $H(s)$
is bounded with bounded derivatives, $\|P^{(n)}(s)\|$ is bounded for all $s$, and hence $P'$ and $P''$ in the constant $C$ [Eq.~\eqref{eq:const_bound}] do not introduce any singularities. 
In addition, since $\mathcal{L}$ is normal, $\| S \| =c/\Delta $ where the constant $c$ depends only on the norm used \cite{kato_perturbation_1995}. Moreover, the identity $S'=S^2 \mathcal{L}' P + P \mathcal{L}' S^2 - S \mathcal{L}' S$ (Appendix \ref{app:identity}) 
implies that $\|S'\| \le 3 \|S\|^2 \|\mathcal{L}'\|$. 
\emph{Thus, from} Eq.~\eqref{eq:const_bound},
\emph{for thermal baths the dependence on the Liouvillian gap is} $C=O(\Delta_{\min}^{-2})$.  
Note that in the absence of the KMS condition one only has $C=O(\Delta_{\min}^{-3})$, implying $T=O(\Delta_{\min}^{-3})$ as a criterion for adiabaticity in accordance with the closed system result of Ref.~\cite{Jansen:07}. 
However, in particular cases the dependence can be even milder. For example, if $\mathcal{L}(s)$ is a unitary family, i.e., $\mathcal{L}(s)= e^{s \mathcal{K}} \mathcal{L}(0) e^{-s \mathcal{K}} $, with $ \mathcal{K}$ an anti-hermitian superoperator, one has  (Appendix \ref{sec:gap_unitary})
 $P' =  \mathcal{K} P -P \mathcal{K} $, so neither $P'$ (nor $P''$) depends on $\Delta_{\min}$.  Moreover $S' =  \mathcal{K}S -S\mathcal{K}$ so that in this case $C=O(\Delta_{\min}^{-1})$, as also shown and extensively exploited in \cite{zanardi_coherent_2014,zanardi_geometry_2015}.

In addition, for thermal baths the constants appearing in Eq.~\eqref{eq:const_bound} bear
an explicit dependence on $H(s)$. E.g., one can show that $\|\rho_{G}{'}(s)\|_{1}\le 2\beta\sqrt{\langle[H'(s)]^{2}\rangle_{G}}$,
where $\langle\bullet\rangle_{G}=\tr[\rho_{G}\bullet]$ is the thermal average (see Appendix \ref{sec:bound_dP}). 
This fact has important consequences for adiabatic quantum computation
where the complexity of a computation is encoded into $H(s)$
and depends on the system size $L$. In general we expect $\| P^{(n)}(s)\| $
to display a stronger divergence with $L$, for some $n$, e.g., at (positive temperature) phase transition points of $H(s)$ %
\footnote{In general one can directly relate $\protect \|P^{(n)}(s)\protect\|$ to expectation values of powers of the Hamiltonian and its derivatives.}. 

When the gap $\Delta_{\min}$ is very small and is attained inside the interval $[0,s]$, the constant $C$ is dominated by the third term in Eq.~(\ref{eq:const_bound}), i.e., $C \simeq \sup_{\sigma} \| S' (\sigma) P'(\sigma) \|$. Using the above estimates for $S'$ and $P'$ we obtain $C \lesssim 6 c^2 \beta \| \mathcal{L}' (\sigma)\|_{\mathrm{max}}\sqrt{\langle[H'(\sigma)]^{2}\rangle_{G,\mathrm{max}}} \Delta_{\min}^{-2}$, where the subscript ``max'' means that the corresponding quantities must be maximized over $\sigma \in [0,s]$. In other words, taking $ T \gtrsim c^2 \beta \| \mathcal{L}' (\sigma)\|_{\mathrm{max}}\sqrt{\langle[H'(\sigma)]^{2}\rangle_{G,\mathrm{max}}} \Delta_{\min}^{-2} /\epsilon$ guarantees adiabaticity up to an error $O(\epsilon)$ in trace norm. 

\section{Case of level crossings}

The gapped case is typical since a random Lindbladian will have a gap above zero with probability one
for all values of $s$. However, symmetries may give rise to degeneracies, and so we would like to extend our result and
consider the case where a finite number of level crossings with the zero eigenvalue may take place along the path
\footnote{A generalization where the spectrum becomes continuous at some point
(e.g., at a second order quantum phase transition) is possible in the unitary case \cite{avron_adiabatic_1999}. Such an extension is not considered here.%
}. 

Since singularities are only algebraic in the finite dimensional case, it is reasonable to expect that, in the case of level crossing, one has $ \| \left[\mathcal{E}(s)-V(s)\right]P(0)\| \sim 1/T^{\eta}$ for large $T$, with a positive exponent $\eta<1$. We are interested in estimating $\eta$ for large $T$. For definiteness assume that at the level crossings the gap vanishes as $\Delta_{\min}(s) \simeq v_i (s-s_i^\ast)^{\alpha_i}$ with some positive exponents $\alpha_i$.  The analysis 
is detailed in Appendix \ref{sec:level_crossing}. 
The final result is
\begin{equation}
\| \left[\mathcal{E}(s)-V(s)\right]P(0)\| \le\sum_{i=1}^{N}\frac{D_{i}}{T^{\eta_{i}}}\ , \qquad \eta_i = \frac{1}{1+\alpha_i}\ ,
\label{eq:eta_bound}
\end{equation}
where $D_i$ are positive constants. Clearly the asymptotic behavior of the right hand side is dictated by the smallest exponent
$\eta_{i}$, i.e., by the largest $\alpha_{i}$, and hence the most divergent of the $N$ gaps.

\begin{figure}[t]
\subfigure[]{ \includegraphics[width=0.48\columnwidth]{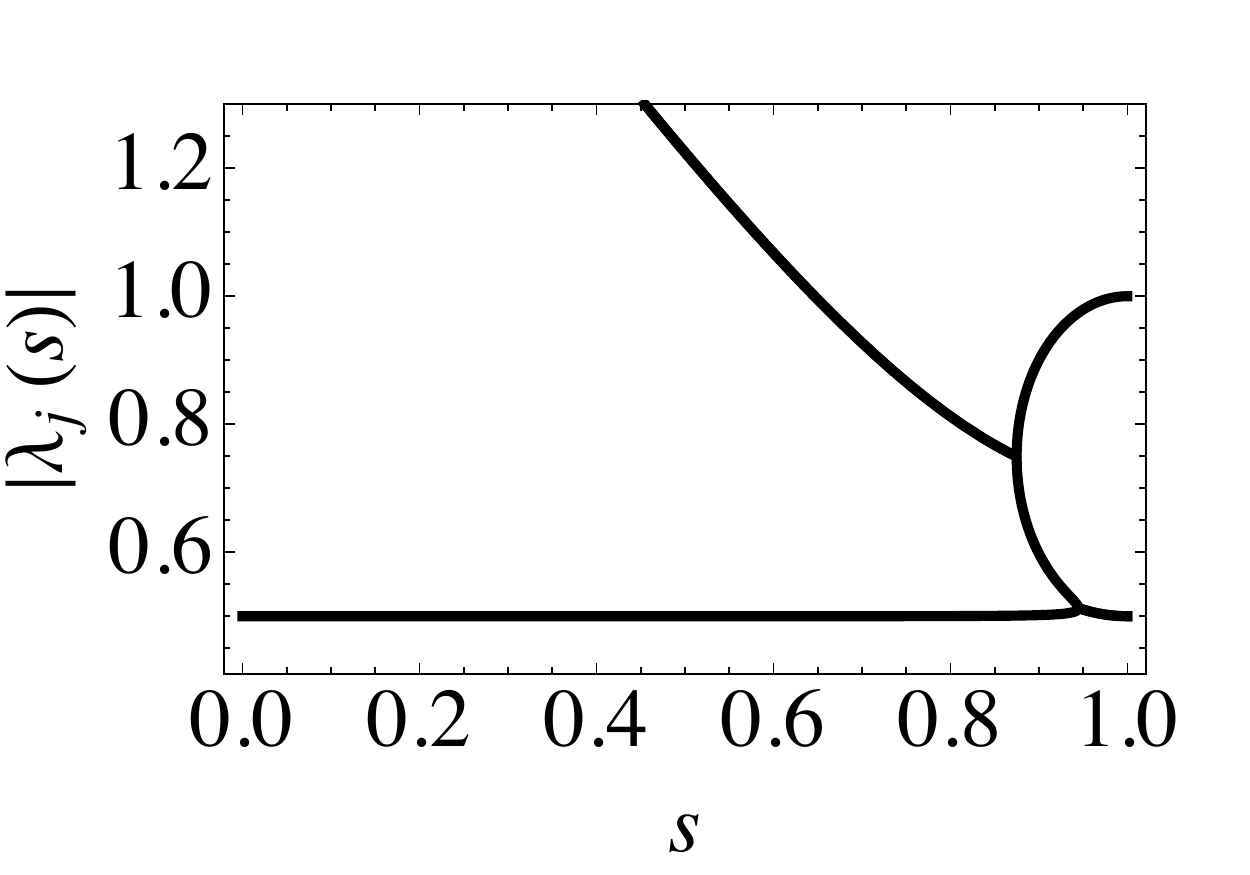}\label{fig:1a}}
\subfigure[]{ \includegraphics[width=0.48\columnwidth]{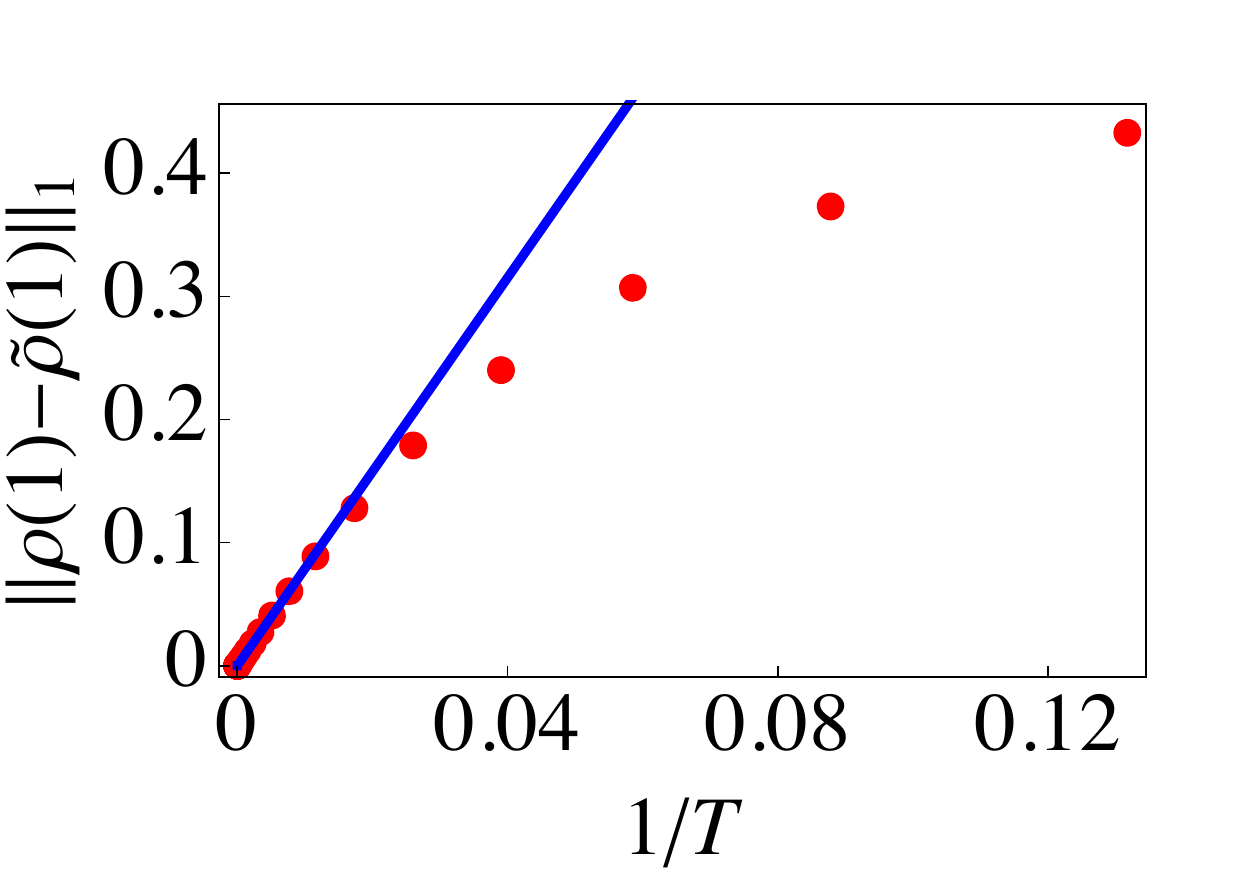}\label{fig:1b}}
\caption{QAT for the gapped case using the model of Example 1. (a) Absolute values
of the eigenvalues of the Lindbladian (zero is not shown). The eigenvalues show square-root singularities at $s\simeq0.88$ and $s\simeq0.94$. (b) Trace norm distance between the actual state and
the ISS for increasing $T$. The blue line is given by 
$\left\Vert \rho(1)-\tilde{\rho}(1)\right\Vert \sim7.88/T$.
Parameters: $m_{x}(s)=1-s$, $m_y(s) = 0$, $m_{z}(s)=s/150$, $\gamma=1/2$ arb.  units Initial condition is $\tilde{\rho}(0)$ as given in Eq.~\eqref{eq:rhoISS1}.}
\label{fig:amplitude_damping}
\end{figure}

\section{Examples}
We now illustrate our results with a few examples. 

\subsection{Example 1}
Let us first consider a time-dependent
generalization of the amplitude damping master equation. The Lindbladian
is $\mathcal{L}(s)=\mathcal{K}(s)+\mathcal{L}_{0}$ with $\mathcal{K}(s)=-i[H(s),\bullet]$,
$H(s)=\boldsymbol{m}(s)\cdot\boldsymbol{\sigma}$ and $\mathcal{L}_{0}=2\gamma\left[\sigma^{-}\bullet\sigma^{+}-(1/2)\left\{ \sigma^{+}\sigma^{-},\bullet\right\} \right]$ ($\sigma^\alpha$ Pauli matrices and $\sigma^{\pm}=\sigma^x\pm i\sigma^y$).
The steady state manifold is one-dimensional. The ISS is given by the solution of $\mathcal{L}(s)\tilde{\rho}(s)=0$ and is (in the $\sigma^z$ basis)
\beq
\tilde{\rho}=\frac{1}{c}\left(\begin{array}{cc}
\boldsymbol{m}^{2}-m_{z}^{2} & -m_{-}(2m_{z}+i\gamma)\\
-m_{+}(2m_{z}-i\gamma) & \boldsymbol{m}^{2}+3m_{z}^{2}+\gamma^{2}
\end{array}\right)\ ,
\label{eq:rhoISS1}
\eeq
where $c=2(\boldsymbol{m}^{2}+m_{z}^{2})+\gamma^{2}$ and $m_{\pm}=m_{x}\pm i m_y$. A plot of the absolute values of the Lindbladian's eigenvalues is shown in Fig.~\ref{fig:1a}. Parameters are chosen to illustrate a phenomenon which
is not possible in the unitary case. Namely, the eigenvalues can have
algebraic singularities of the form $(s-s^{\ast})^{r}$ with $r$
a non-integer rational number. Also shown, in Fig.~\ref{fig:1b}, is the expected scaling of $\| \rho(1)-\tilde{\rho}(1)\|_1$ as $C/T$.

\subsection{Example 2}
Consider a qubit with system Hamiltonian $H(s) = \boldsymbol{m}(s)\cdot\boldsymbol{\sigma}$
interacting with a heat bath at inverse temperature $\beta$, so that the
total Hamiltonian is $H_{\mathrm{tot}}(t)=H(t)+H_{\mathrm{int}}+H_{\mathrm{B}}$, with $H_{\mathrm{int}}=A\otimes B$, where $A$ ($B$) is a system (bath) operator, and $H_{\mathrm{B}}$ is the bath Hamiltonian. {For $H(s)$ of the form of Landau-Zener driving, exact expressions for the transition probabilities for a particular zero-temperature bosonic bath have been obtained in Ref.~\cite{PhysRevLett.97.200404}. To treat the more general case we use the time-dependent Lindblad master equation approximation for a slowly varying system Hamiltonian \cite{ABLZ:12-SI}:} $\mathcal{L}_{\mathrm{tot}}(t)=\mathcal{K}(t)+\mathcal{L}(t)$,
where $\mathcal{K}(t)=-i\left[H(t)+H_{\textrm{LS}}(t),\bullet\right]$.

The dissipative part, specialized to the single qubit case, reads 
\beq
\mathcal{L}(t)=\sum_{\omega}\gamma(\omega)\Big[A_{\omega}(t)\bullet A_{\omega}(t)^{\dagger}-\frac{1}{2}\left\{ A_{\omega}(t)^{\dagger}A_{\omega}(t),\bullet\right\} \Big],
\eeq
with rates $\gamma(\omega)=\int_{-\infty}^{\infty} d \tau e^{i\tau\omega}\langle e^{i\tau H_{\mathrm{B}}} Be^{-i\tau H_{\mathrm{B}}}B\rangle$.
The Lindblad operators are given by the Fourier resolution 
$e^{i\tau H(t)}A e^{-i\tau H(t)}  =  \sum_{\omega}e^{i\omega\tau}A_{\omega}(t)$, where $\omega$ are the Bohr frequencies of $H(t)$. $H_{\textrm{LS}}(t)= \sum_{\omega}S(\omega)A_{\omega}(t)^{\dagger}A_{\omega}(t)$ is the Lamb shift Hamiltonian, with $S(\omega) =  \int_{-\infty}^{\infty}d\omega'\gamma(\omega')\mathcal{P}\left(\frac{1}{\omega-\omega'}\right)$, where $\mathcal{P}$ denotes the principal value.
Moreover, $\gamma_{\alpha}(-\omega)=e^{-\beta\omega}\gamma_{\alpha}(\omega)$
as a consequence of the KMS condition. This implies that the generator
$\mathcal{L}(t)$ together with $\mathcal{K}(t)$ and $\mathcal{K}_{\mathrm{LS}}(t)=-i\left[H_{\mathrm{LS}}(t),\bullet\right]$ all commute at the same time.  

The corresponding Lindbladian has the
following instantaneous eigenvalues (derived in Appendix \ref{app:ex2-deriv}): 
\beq
\lambda = \left\{0, -\gamma(\delta)\left|A_{01}\right|^{2}(1+e^{-\beta\delta}), -\Gamma\pm i\mu\right\} \ ,
\label{eq:eigenvalues}
\eeq
with $2\Gamma=\gamma(0)(\left|A_{00}\right|^{2}+\left|A_{11}\right|^{2})+\gamma(\delta)\left|A_{01}\right|^{2}(1+e^{-\beta\delta})$
and $\mu=\delta-S(0)(\left|A_{00}\right|^{2}-\left|A_{11}\right|^{2})+\left|A_{01}\right|^{2}[S(\delta)-S(-\delta)]$,
where $\delta=2\| \boldsymbol{m}\| $ is the instantaneous Hamiltonian
gap, $A_{ab}=\bra{\epsilon_a}A\ket{\epsilon_b}$, and $|\epsilon_a\rangle$, $a=\{0,1\}$  denotes
the instantaneous eigenvectors of $H(s)$.

\begin{figure}
\subfigure[]{ \includegraphics[width=0.48\columnwidth]{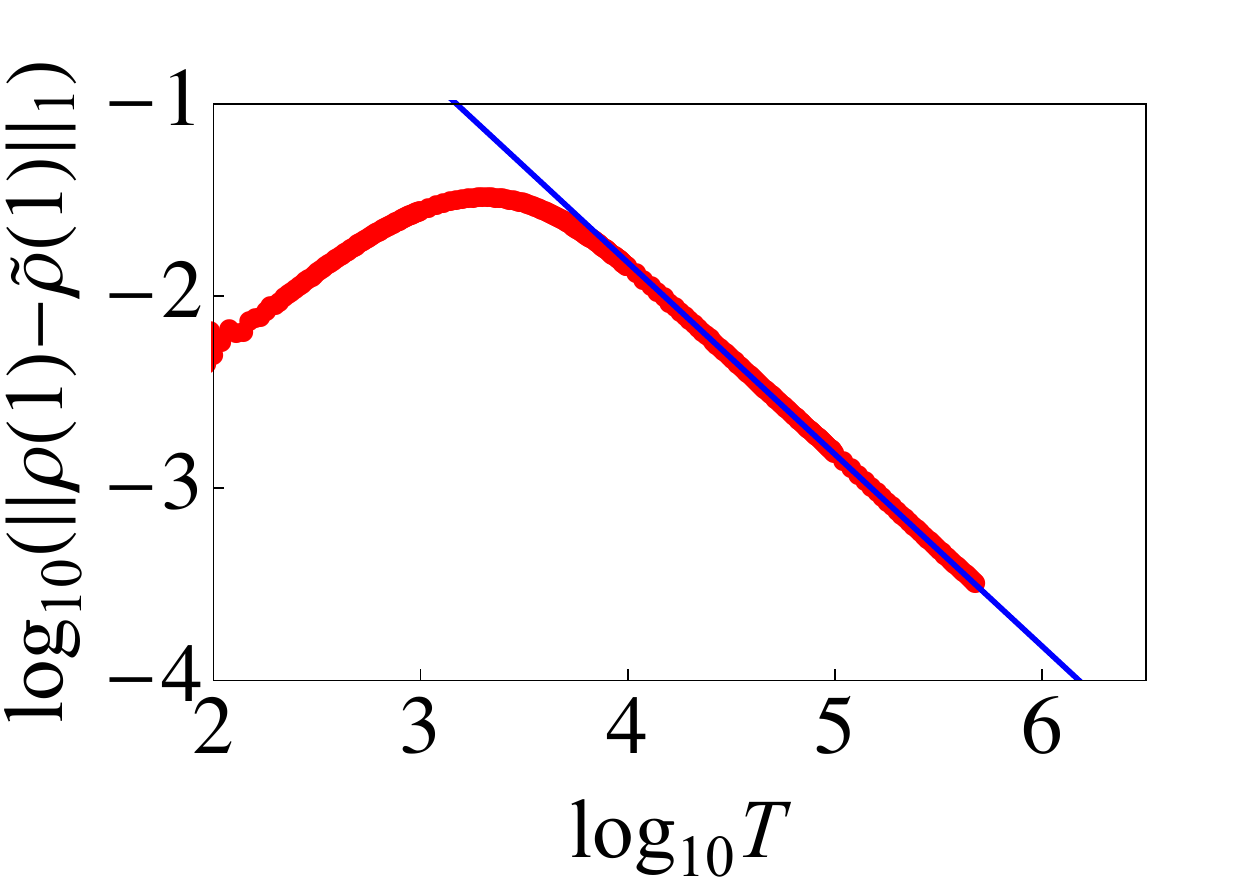}\label{fig:3a}}
\subfigure[]{ \includegraphics[width=0.48\columnwidth]{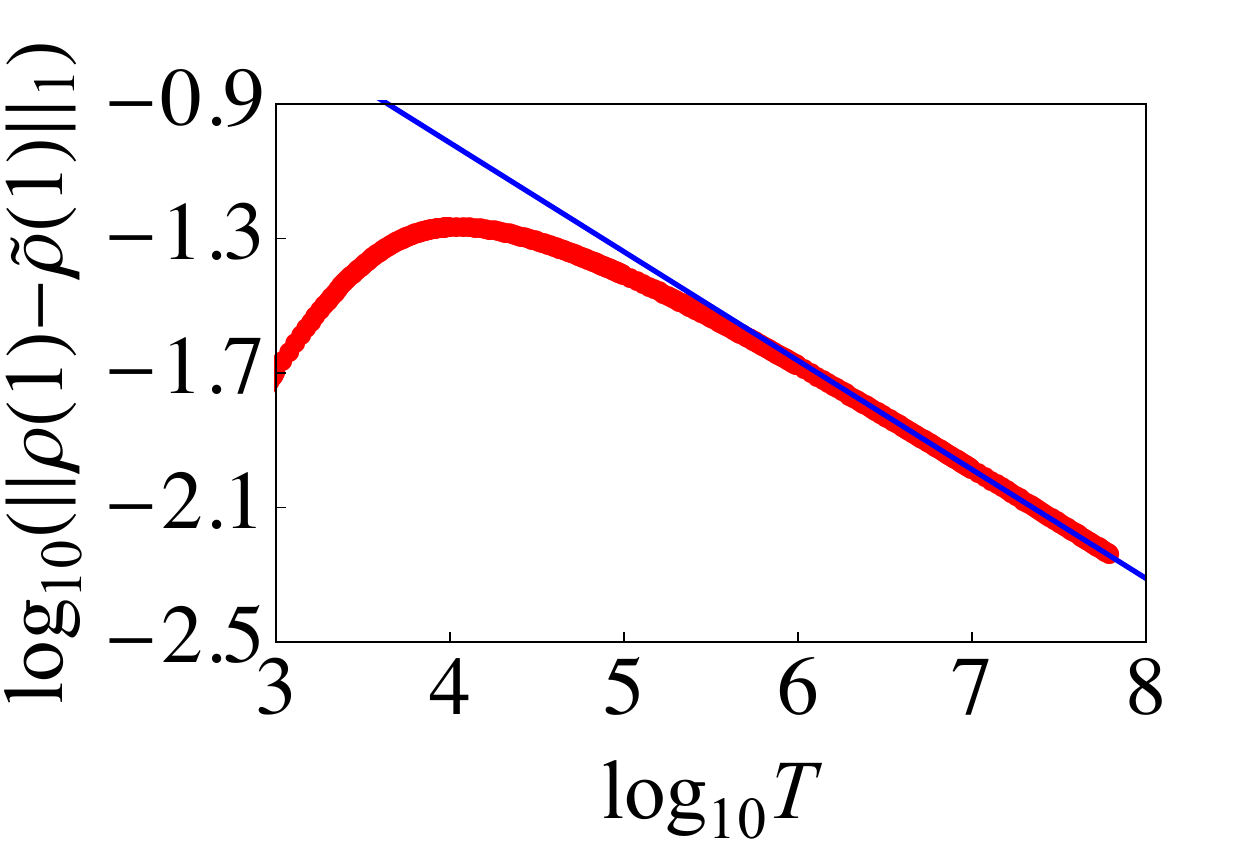}\label{fig:3b}}
\caption{Illustration of the QAT for a single qubit coupled to a thermal bath (Example 2). (a)
Gapped case with $A=g\sigma^{y}$. The fit (blue line) gives $\| \rho(1)-\tilde{\rho}(1)\|_1 = 148.5/T^{0.9990}$.
(b) Gapless case with $A=g\sigma^{z}$. Now the fit 
gives $\| \rho(1)-\tilde{\rho}(1)\|_1 = 1.910/T^{0.324}$. Fits are for $T\geq 10^{5.4}$.  Parameters: $g = 10^{-2}$, $\omega_x = \omega_z = -1/2$ arb.  units, $\beta = 1$ arb.  units, $\gamma(\omega) = \frac{2 \pi \omega e^{-|\omega|/8\pi}}{1 - e^{- \beta \omega}}$.}
\label{fig:tameem1}
\end{figure}

We now illustrate the QAT for $H(s)=\omega_{x}(1-s)\sigma^{x}+\omega_{z}s\sigma^{z}$
and system operator $A=g\sigma^{y,z}$, where $g$ is a coupling
constant. For $A=g\sigma^{y}$, there is always a gap in the Liouvillian
spectrum above the zero eigenvalue. Correspondingly the decay is $T^{-1}$
as shown in Fig.~\ref{fig:3a}. For $A=g\sigma^{z}$ 
the system Hamiltonian commutes with $A$ when $s=s^\ast=1$, at which point the spectrum becomes degenerate. Indeed one can check that $\left|A_{01}\right|^{2}=\omega_{x}^{2}(1-s)^{2}/\| \boldsymbol{m}(s)\| ^{2}$
and, correspondingly, from Eq.~\eqref{eq:eigenvalues}, the second eigenvalue
goes to zero quadratically in $s-s^\ast$: $\lambda_{2}(s)\simeq-\gamma(\delta_{})(\omega_{x}/\omega_{z})^{2}(1-s)^{2}(1+e^{-\beta\delta_{}})$.
This corresponds to $\alpha=2$ and hence [recall Eq.~\eqref{eq:eta_bound}] an exponent $\eta=1/3$. Our numerical simulations are in agreement with this prediction, as seen in Fig.~\ref{fig:3b}. Note that the Hamiltonian gap $\delta$ enters only indirectly via $\lambda$.

\section{Conclusions}

Using a novel adiabatic expansion we have extended the adiabatic theorem of quantum mechanics to
open systems described by a time-dependent master equation with generator
$\mathcal{L}(t)$ in Lindblad form. The theorem, first proven using different methods in \cite{joye_general_2007,salem_quasi-static_2007,Avron:2012tv}, states that if one initializes the system
in $\mathrm{Ker}\mathcal{L}(0)$ and a gap condition is satisfied, the evolution brings the system close to
$\mathrm{Ker}\mathcal{L}(T)$ up to an error $C/T$, where $C$ is a constant and $T$ the total time. Our approach allowed us to extend the results of \cite{joye_general_2007,salem_quasi-static_2007,Avron:2012tv} in two directions particularly relevant for quantum state preparation and quantum annealing in open systems. On the one hand, we related the constant $C$ to the smallest (in absolute value) gap  $\Delta_{\min}$ of the Liouvillian. For general Liouvillians we obtained $C=O(\Delta^{-3}_{\min})$, whereas for thermal baths satisfying the KMS condition we found an improved scaling $C=O(\Delta^{-2}_{\min})$.  More precisely, we showed that taking
$ T \gtrsim \beta \| \mathcal{L}' (\sigma)\|_{\mathrm{max}}\sqrt{\langle[H'(\sigma)]^{2}\rangle_{G,{\mathrm{max}}}} \Delta_{\min}^{-2} /\epsilon$, guarantees adiabaticity up to an error $O(\epsilon)$ in trace norm.
On the other hand, we extended previous results to the case of level crossing, for which the
error becomes $O(T^{-\eta})$ with an exponent $\eta$ that depends on the rate of the gap closing. Thus  level crossings with the instantaneous steady state can slow convergence down. 
We provided several examples to illustrate our findings, which confirm the predicted scaling with $T$. 
An interesting open question is whether the growing body of techniques developed for bath engineering \cite{diehl_quantum_2008,verstraete_quantum_2009,zanardi_dissipative_2015} can be used to enact  boundary cancellation methods and reduce the error to $O\left(T^{-n}\right)$ with controllable $n>1$, as in the closed system case \cite{lidar:102106,RPL:10,Wiebe:12,Ge:2015wo}.
Our results have implications for adiabatic quantum computation and quantum annealing in the presence of dissipation, where the closed-system adiabatic theorem cannot be directly applied. 

\acknowledgments
This work was supported under under ARO MURI Grant No. W911NF-11-1-0268 and ARO grant number W911NF-12-1-0523.

\appendix

\section{Proof that zero is a semi-simple eigenvalue of $\mathcal{L}(s)$}

\label{app:0} 
We assume in this section that $\mathcal{L}(s)$ is finite-dimensional. We show that zero
is a semi-simple eigenvalue of $\mathcal{L}(s)$, i.e., that there
are no idempotents in the Jordan block decomposition of $\mathcal{L}$
corresponding to its zero eigenvalue. Assume the contrary, i.e., that one can write
$\mathcal{L}(s)= 0\times P(s)+D(s) +R(s)$ where $D(s)$ is nilpotent, i.e., $D(s)^m=0$, $D(s)^{m-1}\neq 0$, and $R(s)P(s)=P(s)R(s)=0$. Then
\begin{equation}
e^{t\mathcal{L}(s)}P(s) =  P(s)+\sum_{n=1}^{m-1} \frac{t^n D(s)^n}{n!} \label{eq:semisimple}
\end{equation}
where we also used that $D(s)P(s)=P(s)D(s)=D(s)$. By assumption, $\mathcal{L}(s)$ is Markovian, which means that for each $s$ and $t>0$, $e^{t\mathcal{L}(s)}$ is CPTP and so $\|e^{t\mathcal{L}(s)}\|=1$.  Taking the norm of both sides of Eq.~(\ref{eq:semisimple}) we obtain that the left hand side is bounded while the right hand side grows unboundedly with $t$, which is a contradiction. Hence $D(s)=0$ and  
$\mathcal{L}(s)P(s)=P(s)\mathcal{L}(s)=0$. For an alternative proof see, e.g., Ref.~\cite{wolf_quantum_2012} or Proposition 5 of Ref.~\cite{Avron:2012tv}.  

Note that when the generator $\mathcal{L}(s)$ is normal, (i.e., $[\mathcal{L}(s),\mathcal{L}^{\ast}(s)]=0$, where $\ast$ denotes the adjoint with respect to the scalar product  $\langle X,Y\rangle_{G}:=\tr[\rho_{G}(s)X^{\dagger}Y]$), as is the case for a thermal bath, this has the pleasant consequence that all the eigenprojectors of $\mathcal{L}(s)$ are bounded (in fact have norm one) and there are no nilpotent terms in its Jordan block decomposition.

\section{Proof that $P(s)V(s)=V(s)P(0)$}
\label{app:intertwiner} 
Recall that we defined the intertwiner to
be the solution of the following differential equation: $V'(s)=[P'(s),P(s)]V(s)$.
To check that $V(s)$ satisfies the intertwining relation, define $W(s)=P(s)V(s)$.
From now on we drop the $s$ dependence when not explicitly needed.
Differentiating $P^{2}=P$ gives $P'=P'P+PP'$, and after right-multiplying
by $P$ gives $PP'P=0$ and likewise $QQ'Q=0$ [which also implies $QP'Q=0$ since $0=Q\1'Q=Q(P'+Q')Q$]. Thus $[[P',P],P]=P'P-2PP'P+PP'=P'$. Using
this, note that $W'=P'V+PV'=(P'+P[P',P])V=[P',P]PV=[P',P]W$, which
is the same differential equation as the one satisfied by $V$. Since
$W$ and $V$ differ in their initial condition, i.e., $V(0)=\1$
and $W(0)=P(0)$, it follows that $W(s)=V(s)P(0)$, i.e., 
\beq
W(s) = P(s)V(s)=V(s)P(0)\ .
\label{eq:B1}
\eeq

\section{Complete positivity of $V(s)P(0)$, and lack of positivity of $V(s)$}

\label{app:positivity}

Here we show that $W(s)          =V(s)P(0)$ is a CPTP map even though the intertwiner
$V(s)$ itself is not in general. From Appendix~\ref{app:intertwiner} we obtain
\begin{align}
W' & =[P',P]W\notag\\
 & =P'PW-PP'W=P'PW-PP'PW\notag\\
 & =P'PW=P'W\ .\label{eq:W'}
\end{align}
Now we write the solution of the differential equation using the method
of Euler lines \cite{Euler:book,Butcher:book}, i.e., $W(s)=\lim_{N\to\infty}\left(\1+\epsilon P'(s-\epsilon)\right)\cdots\left(\1+\epsilon P'(0)\right)P(0)$,
with $\epsilon=s/N$. Using $P(s)^{2}=P(s)$ we get, in the $\epsilon\rightarrow0$
limit, 
\[
\left(\1+\epsilon P'(0)\right)P(0)=\left(P(0)+\epsilon P'(0)\right)P(0)=P(\epsilon)P(0)\ ,
\]
where the second equality is up to first order in $\epsilon$. Then (with the same notation) 
\[
\left(\1+\epsilon P'(\epsilon)\right)P(\epsilon)=\left(P(\epsilon)+\epsilon P'(\epsilon)\right)P(\epsilon)=P(2\epsilon)P(\epsilon)\ ,
\]
etc., until $W(s)=\lim_{N\to\infty}P(s)\cdots P(2\epsilon)P(\epsilon)P(0)$.
In this form, $W(s)$ is an infinite product of CPTP maps, so it
is a CPTP map itself. It follows from submultiplicativity of the norm \cite{Bhatia:book}
that $\|W(s)\| = \|V(s)P(0)\|\leq1$.

Next, let us demonstrate that $V(s)$, in general, is not even positive. Consider, e.g., the case where $P(s)$ is one-dimensional, so that $P(s) = |\rho(s) \rangle \langle \1 |$, where $\langle x|$ is the adjoint of $ |x\rangle$ with respect to the Hilbert-Schmidt scalar product $\braket{A}{B} = \tr[A^\dagger B]$.
We assume that $\rho(s)$ is differentiable. Then $P'(s) = |\rho'(s) \rangle \langle \1 |$ and so  $P' P = P'$ while $P P' = 0$ since $\rho'(s)$ is traceless. Thus $V'=[P',P]V$ becomes $V'= P' V$ (with $V(0)=\1$), with the solution
\beq
V(s) = \mathrm{Texp} \left ( \int_0^s P'(\sigma) d\sigma \right )\ ,
\eeq
where $ \mathrm{Texp} $ denotes the time-ordered exponential.
However $P'(s_1) P'(s_2) = P'(s_2) P'(s_1) = 0$ [again, since $\rho'(s)$ is traceless] so the time-ordered exponential in the above equation reduces to a standard exponential. The solution is then 
\begin{align}
V(s) &= \exp{ \left [ (|\rho(s) \rangle - |\rho(0) \rangle) \langle \1 | \right ]} \nonumber \\
     &= {\1} + (|\rho(s) \rangle - |\rho(0) \rangle) \langle \1 | \ , 
\end{align}
where in the second line we used the fact that $[(|\rho(s) \rangle - |\rho(0) \rangle)\langle \1 |]^2 =0$. 
Now, it is easy to see that, unless $\rho(s)=\rho(0)$ such that $V(s)=\1$,  $V(s)$ does not even preserve positivity. 
Let us define, for clarity, $\delta \rho:= \rho(s)  - \rho(0)$. This is a hermitian and traceless operator, so it must have a negative eigenvalue, i.e., there exist  a vector $|\alpha\rangle$ and a real number $\alpha >0$, such that $\delta \rho  |\alpha\rangle = -\alpha |\alpha\rangle$. Let us consider the following state 
\begin{equation}
x_0 := \lambda \frac{\1}{d} + (1-\lambda) | \alpha^\perp \rangle \langle \alpha^\perp |,
\end{equation}
where $d$ is the dimension of the ISS, $| \alpha^\perp \rangle $ is a vector satisfying $\langle \alpha | \alpha^\perp \rangle = 0$ and $\lambda \in [0,1]$. It is clear that $x_0$ is a positive operator. However,
\begin{equation}
V(s) x_0 |\alpha\rangle = \left ( \frac{\lambda}{d} -\alpha \right )  | \alpha \rangle \ ,
\end{equation}
such that, for $\lambda < \min (\alpha d,1 )$, $V(s) x_0$ has a negative eigenvalue, i.e., $V(s)$ does not preserve positivity. 
 
\section{Proof of Eq.~(2)}
\label{sec:arb_order}

We assume that $\mathcal{L}$ is $m$-fold differentiable.

\subsection{The $m=1$ case}
The proof of Eq.~(2) 
 of the main text for $m=1$ closely follows the classic
reference \cite{kato_adiabatic_1950}. As above, let $P(s)$ be the
projector onto the instantaneous, zero eigenvalue of $\mathcal{L}(s)$. In this subsection we do not require the ISS to be unique. Since we established in Appendix~\ref{app:0} that $\mathcal{L}P=P\mathcal{L}=0$ (from now on we omit the explicit dependence on $s$ if not strictly needed), it follows that $\mathcal{L}=Q\mathcal{L}=\mathcal{L}Q$. Therefore the reduced resolvent $S=\lim_{z\to0}Q(\mathcal{L}-z)^{-1}Q$ satisfies $S\mathcal{L} = \lim_{z\to0}Q(\mathcal{L}-z)^{-1}Q\mathcal{L} = Q\lim_{z\to0}(\mathcal{L}-z)^{-1}\mathcal{L} =Q$. To summarize: 
\bes 
\begin{align}
\mathcal{L} S  & =S \mathcal{L} =Q \label{eq:S-props1} \ , \\
S P  & =P S =0\ .\label{eq:S-props2}
\end{align}
\ees It follows from Eq.~\eqref{eq:W'} and $PP'P=0$ that $PW'=0$ 
and so 
\beq W'=QW'\ . 
\label{eq:Wprime} 
\eeq

Recall [Eq.~(1) 
of the main text] that the evolution operator $\mathcal{E} $
satisfies $\mathcal{E}' =T\mathcal{L} \mathcal{E} $. Now, $0=(\mathcal{E}\mathcal{E}^{-1})'=\mathcal{E}'\mathcal{E}^{-1}+\mathcal{E}(\mathcal{E}^{-1})'$,
so that
\beq 
(\mathcal{E}^{-1})' = -T\mathcal{E}^{-1}\mathcal{L}\ . 
\label{eq:E-1'}
\eeq 
Therefore $(\mathcal{E}^{-1} )'W =-T\mathcal{E}^{-1} \mathcal{L} P V =0$,
since $\mathcal{L} P =0$. Hence $(\mathcal{E}^{-1}W)'=\mathcal{E}^{-1}W'$, which integrated over $[0,s]$ gives 
\begin{equation}
 \mathcal{E}^{-1}(s)V(s)P(0)-P(0)=\int_{0}^{s}\mathcal{E}^{-1}(\sigma)W'(\sigma)d\sigma \ .
 \label{eq:start}
\end{equation}

Now we use 
\begin{align}
 \mathcal{E}^{-1}W' & =  \mathcal{E}^{-1}Q W'  \nonumber\\  
 & = \mathcal{E}^{-1}\mathcal{L}SW'\nonumber\\ 
 & =-T^{-1}(\mathcal{E}^{-1})'SW'\ , \label{eq:D3c}
\end{align}
where we used Eqs.~\eqref{eq:Wprime},
\eqref{eq:S-props1}, and \eqref{eq:E-1'}.
We plug this into Eq.~(\ref{eq:start}) and integrate by parts, to obtain
\begin{multline}
  \mathcal{E}^{-1}(s)W(s)-P(0) =\\
  \frac{1}{T}\left\{ \int_{0}^{s}\mathcal{E}^{-1}(\sigma)[S(\sigma)W'(\sigma)]'d\sigma-\left.\left(\mathcal{E}^{-1}SW'\right)\right|_{0}^{s}\right\} . \label{eq:X1}
\end{multline}
Now we act with $\mathcal{E}(s)$
from the left and use the property $\mathcal{E}(a,b)\mathcal{E}(b,0)=\mathcal{E}(a,0)$ to obtain
\begin{multline}
 [\mathcal{E}(s)-V(s)]P(0) =
  \frac{1}{T}  \left.\mathcal{E}(s,\sigma)S(\sigma)W'(\sigma)\right|_{0}^{s} \\
  -\frac{1}{T} \int_{0}^{s}d\sigma\,\mathcal{E}(s,\sigma)  \left(S'W' +SW''\right)(\sigma) \ .  \label{eq:D6}
\end{multline}
At this point recall that $W'=P'W$ [Eq.~\eqref{eq:W'}] and hence [using Eq.~\eqref{eq:B1}] $W'(s)=P'(s)V(s)P(0)$, and also $W''=[P''+(P')^2]W$. It follows from the latter, together with $S=SQ$ [from Eq.~\eqref{eq:S-props2}] and $W=PW$ [from Eq.~\eqref{eq:B1}], that $SW'' = SQ [P''+(P')^2]P W$. But $Q (P')^2 P =QP'(P+Q)P'P=0$ because $PP'P=0$ and $Q P' Q=0$ (see Appendix~\ref{app:intertwiner}), so that $SW'' = SP'' W$. Collecting all these results we finally obtain
\begin{multline}
[\mathcal{E}(s)-V(s)]P(0)=\\
\frac{1}{T}\Big\{\left(S(s)P'(s)V(s)P(0)-\mathcal{E}(s)S(0)P'(0)P(0)\right)\\
-\int_{0}^{s}d\sigma\,\mathcal{E}(s,\sigma)\left[S'P'+SP''\right](\sigma)V(\sigma)P(0)\Big\}\ .
\label{eq:D7}
\end{multline}
Using submultiplicativity along with $\|P(s)\|=1$ and $\|V(s)P(0)\|\leq1$ (as shown in Appendix~\ref{app:positivity}) and $\|\mathcal{E}(s)\|\leq1$ (see immediately below) we obtain 
\begin{equation}
\| \left[\mathcal{E}(s)-V(s)\right]P(0)\| \le C/T\ ,
\end{equation}
with
\begin{multline}
C=\| S(s)\| \| P'(s)\| +\| S(0)\| \| P'(0)\| \\
+\sup_{\sigma\in[0,s]}\| [S'P'+SP''](\sigma)\| \ , \label{eq:constC}
\end{multline}
as stated in Eqs.~(3) and (4)
 of the main text.

Note that when restricted to acting on normalized states, clearly $\protect\|\mathcal{E}\protect\|=1$
by trace preservation. More generally, since $\mathcal{L}(s)$ is
a Lindbladian we have $\protect\|\exp[t\mathcal{L}(s)]\protect\|\leq1$ for each fixed
$s$ and $t>0$ [i.e., $\mathcal{L}(s)$ generates a contraction for any fixed
$s$]. Using the method of Euler lines (see Appendix~\ref{app:positivity})
one can show that this implies that $\protect\|\mathcal{E}(s,t)\protect\|\leq1$,
because $\mathcal{E}(s,t)$ can be written as an infinite product
of evolutions of the form $\exp[t\mathcal{L}(s_{i})]$.

\subsection{Extension to arbitrary order}

Next we show how to extend the integration by parts technique of the previous subsection
to arbitrary order $m$. A particularly simple series will result under the assumption that
the ISS is unique, i.e., $P(s)=|\rho(s)\rangle\langle\1|$, as in Appendix~\ref{app:positivity} (this assumption is not essential, but it significantly simplifies our calculations below). 
As before, we also assume that $\mathcal{L}$
generates a trace preserving map. Both of the latter two assumptions are
usually satisfied for thermal (i.e., Davies) generators. 

We begin by
considering the first term of Eq.~\eqref{eq:X1}, i.e.,
$(SW')'=S'W'+SW''$.
Our strategy will be to repeat the integration by parts on the term arising
from $W'$, while keeping (i.e., not integrating by parts) the term arising from $W''$. We will show how this can be done repeatedly and thus extend the result to arbitrary order.

Now, thanks to the assumption of uniqueness $P(s)=|\rho(s)\rangle\langle\1|$. Hence
from $PS=0$ we obtain $\langle\1|S=0$ and so $P'S=|\rho'(s) \rangle \langle \1 |S=0$. Using $P'S+PS'=0$ we get  $PS'=0$ so that, writing $S'Q = (P+Q)S'Q = PS'Q+QS'Q$ it thus follows that 
\begin{equation}
S'Q=QS'Q\ .
\label{eq:S'Q}
\end{equation}
This means that, using Eq.~\eqref{eq:Wprime}, we can write 
\begin{equation}
S'W'=S'QW'=QS'W'\ . 
\label{eq:S'W'}
\end{equation}

Returning to the term of interest in Eq.~\eqref{eq:X1}, we have
\bes
\label{eq:E2}
\begin{align}
\mathcal{E}^{-1}S'W' & =\mathcal{E}^{-1}QS'W' \label{eq:E2a}\\
 & =\mathcal{E}^{-1}\mathcal{L}SS'W' \label{eq:E2b}\\
 & =-T^{-1}(\mathcal{E}^{-1})'SS'W' \label{eq:E2c}\ ,
\end{align}
\ees
where in the last equality we used Eq.~\eqref{eq:E-1'}. This term can now be integrated by parts again, in analogy to the term in Eq.~\eqref{eq:D3c}. 
Differentiating $SS'W'$, we
obtain two terms, one of which is $SS'W''$, which we keep. The other is $(SS')'W'$
which we would like to integrate by parts again (because it diverges more strongly when $\Delta_{\mathrm{min}} \to 0$). We now show that this
procedure can be iterated to any order provided $\mathcal{L}$ is
differentiable sufficiently many times. 

Let us assume that at order $n$ we obtained the term 
\begin{equation}
\frac{1}{T^{n}}\int_{0}^{s}\!d\sigma\,\mathcal{E}^{-1}X_{n}'W'.
\end{equation}
Assume for the moment that $X_{n}'W'=QX_{n}'W'$; we will show shortly that this is legitimate.
In this case we can repeat the calculation of Eq.~\eqref{eq:E2} and write $\mathcal{E}^{-1}X_{n}'W' = -T^{-1}(\mathcal{E}^{-1})'SX_{n}'W'$. We can thus use the integration by parts trick:
\begin{align}
&\frac{1}{T^{n}}\int_{0}^{s}\!d\sigma\,\mathcal{E}^{-1}X_{n}'W' =-\frac{1}{T^{n+1}}\int_{0}^{s}\!d\sigma\,(\mathcal{E}^{-1})'SX_{n}'W' \nonumber\\
 & =-\frac{1}{T^{n+1}}\left.\mathcal{E}^{-1}SX_{n}'W'\right|_{0}^{s}+\frac{1}{T^{n+1}}\int_{0}^{s}\!d\sigma\,\mathcal{E}^{-1}(SX_{n}'W')' \nonumber\\
 & =-\frac{1}{T^{n+1}}\left.\mathcal{E}^{-1}SX_{n}'W'\right|_{0}^{s}+\frac{1}{T^{n+1}}\int_{0}^{s}\!d\sigma\,\mathcal{E}^{-1}SX_{n}'W''\nonumber\\
 & \quad+\frac{1}{T^{n+1}}\int_{0}^{s}\!d\sigma\,\mathcal{E}^{-1}(SX_{n}')'W'.
\end{align}
The first integral (with $W''$) is the one we keep, while the second (with $W'$) is the one we continue to integrate by parts. This shows that the process can be iterated with $X_{n+1}=SX_{n}'$. 

We now verify that we can plug in a $Q$ term at each order as claimed above. 
We prove it by induction.
Assume that $ X_{n}'W'=Q X_{n}'W'$. We
wish to show that this implies that $X_{n+1}'W'=QX_{n+1}'W'$.
But this is clear since 
\begin{align}
& X_{n+1}'W'  =(S'X_{n}'+SX_{n}'')W' \notag \\
& =(S'QX_{n}'+QSX_{n}'')QW' = Q(S'QX_n'+SX''_n)QW' \notag \\
&= Q(S'X_n'+SX''_n)W' = QX_{n+1}'W'\ ,
\end{align}
where we used the induction hypothesis, Eq.~\eqref{eq:Wprime} and Eq.~\eqref{eq:S'Q}. We have already shown that the claim holds for $n=1$ [Eq.~\eqref{eq:S'W'} for $X_1=S$] so we are done.  

We thus obtain
\begin{align}
\mathcal{E}^{-1}(s)W(s)-P(0) & =\sum_{n=1}^{m}\frac{\Gamma_{n}}{T^{n}}+\frac{1}{T^{m}}\int_{0}^{s}\!d\sigma\,\mathcal{E}^{-1}X_{m}'W'\\
\Gamma_{n} & =-\left.\mathcal{E}^{-1}X_{n}W'\right|_{0}^{s}+\int_{0}^{s}\!d\sigma\,\mathcal{E}^{-1}X_{n}W''\\
X_{n+1} & =S X_{n}'\ ,\quad X_{1}=S.
\end{align}
Multiplying from the left by $\mathcal{E}(s)$ we obtain 
\begin{align}
[\mathcal{E}(s)-V(s)]P(0) & =\sum_{n=1}^{m}\frac{\Omega_{n}}{T^{n}} \nonumber \\
& -\frac{1}{T^{m}}\int_{0}^{s}\!d\sigma\,\mathcal{E}(s,\sigma)X_{m}'(\sigma)W''(\sigma)\\
\Omega_{n} & =\left.\mathcal{E}(s,\sigma)X_{n}(\sigma)W'(\sigma)\right|_{0}^{s}  \nonumber \\
& -\int_{0}^{s}\!d\sigma\,\mathcal{E}(s,\sigma)X_{n}(\sigma)W''(\sigma). 
\end{align}
If $\mathcal{L}$ is differentiable infinitely many times, we can
write 
\begin{equation}
[\mathcal{E}(s)-V(s)]P(0)=\sum_{n=1}^{\infty}\frac{\Omega_{n}}{T^{n}}\ ,
\end{equation}
under the assumption of convergence. We expect this assumption
to hold in the finite-dimensional case if $\mathcal{L}$
depends analytically on $s$. 

\section{Additional remarks on the closed system limit}
\label{app:closed}

The closed system limit may also be achieved by considering ${\cal L}_x(s)= {\cal K}(s)+x{\cal D}(s)$, where ${\cal D}(s)$ is a dissipative generator, and taking the (weak coupling) limit $x\to 0$. This is a singular limit: for $x\neq 0$ there is typically only one steady state, whereas for $x=0$ there is a large degeneracy ($\ge d$ Hilbert's space dimension). Calling  $P(x)$ the spectral projection onto the zero eigenvalue for fixed $s$, one has  $\lim_{x\to 0 } P(x) \neq P(0)$, since the left hand side has rank one, whereas $P(0)$ has rank $\ge d$. Moreover,  for sufficiently small $x$ the smallest gap will be $\propto x$ and hence also $(1/{\rm gap})^\eta$ ($\eta>0$) will diverge. Thus the constant $C$ in Eq.~\eqref{eq:const_bound}, and even the optimal constant $C_{\mathrm{opt}}$ [obtained taking the infimum of all the constants $C$ satisfying the bound Eq.~\eqref{eq:adia_bound} for all $T$] will, in general, likely be large. In other words, in general the limit $x\to 0$ is singular and the adiabatic theorem for closed system cannot  be recovered in this form. A possible way to recover it is to abandon projectors and move to states, at least in the case of non-degenerate ISS for $x\neq0$. In other words, let us call $\rho_{x}(s)$ the state evolved with generator ${\cal L}_x(s)$ from $\sigma_{x}(0)$ [${\cal L}_x(s)\sigma_{x}(s) = 0$ is assumed to be unique for $x\neq0$]. Now it is possible that $\rho_{x}(s)$ and $\sigma_{x}(s)$ have a well defined limit as $x\to 0$ such that the following bound
\begin{equation}
\| \rho_{x}(s)- \sigma_{x}(s)\| \le \tilde{C}_{x}/T\ 
\end{equation}
may admit a non-trivial limit for $x\to0$ with  $\lim_{x\to0} \tilde{C}_{x} <\infty$. This may be achieved by applying Eq.~\eqref{eq:D7} to $\sigma_{x}(0)$ and bounding the resulting expression.

\section{Extension to the case of a degenerate kernel}
\label{sec:degenerate_kernel}

If the kernel of $\mathcal{L}(s)$ is not one-dimensional, in general
we will also have a $P\mydash Q$ block in $X_{n}$. The procedure,
however, can still be iterated.  Assume that at order $n$ we obtained
\begin{multline}
\frac{1}{T^{n}}\int_{0}^{s}d\sigma\,\mathcal{E}^{-1}X_{n}W'=\\
\frac{1}{T^{n}}\int_{0}^{s}d\sigma\,\left[\mathcal{E}^{-1}PX_{n}QW'+\mathcal{E}^{-1}QX_{n}QW'\right]\ .
\end{multline}
We keep the first term (we will have to show later that it scales
nicely), and integrate the second by parts, for which we use $\mathcal{E}^{-1}QX_{n}QW'=\mathcal{E}^{-1}\mathcal{L}SX_{n}QW'=-T^{-1}(\mathcal{E}^{-1})'SX_{n}W'$.
The above equation becomes
\begin{align}
 & \frac{1}{T^{n}}\int_{0}^{s}d\sigma\,\mathcal{E}^{-1}PX_{n}QW'-\frac{1}{T^{n+1}}\mathcal{E}^{-1}SX_{n}W'\Big|_{0}^{s}\nonumber \\
 & +\frac{1}{T^{n+1}}\int_{0}^{s}d\sigma\,\mathcal{E}^{-1}(SX_{n}W')'\\
= & \frac{1}{T^{n}}\int_{0}^{s}d\sigma\,\mathcal{E}^{-1}PX_{n}QW'-\frac{1}{T^{n+1}}\mathcal{E}^{-1}SX_{n}W'\Big|_{0}^{s}\nonumber \\
 & +\frac{1}{T^{n+1}}\int_{0}^{s}d\sigma\,\mathcal{E}^{-1}(SX_{n})W''\nonumber \\
 & +\frac{1}{T^{n+1}}\int_{0}^{s}d\sigma\,\mathcal{E}^{-1}(SX_{n})'W'\ ,
\end{align}
from which we see that the recurrence is $X_{n+1}=(SX_{n})'$ with
$X_{0}=\1$ as previously. However, now the series reads
\begin{align}
\mathcal{E}^{-1}(s)W(s)-P(0) & =\sum_{n=1}^{\infty}\frac{\Gamma_{n}}{T^{n}}\\
\Gamma_{n} & =-\left.\mathcal{E}^{-1}X_{n}W'\right|_{0}^{s}\nonumber \\
 & +\int_{0}^{s}d\sigma\,\mathcal{E}^{-1}X_{n}W''\nonumber \\
 & +\int_{0}^{s}d\sigma\,\mathcal{E}^{-1}PX_{n}'QW'\ ,
\end{align}
with $X_{n+1}=S(X_{n}'),\quad X_{1}=S$. Multiplying from the left
by $\mathcal{E}(s)$ we obtain
\begin{align}
\left[\mathcal{E}(s)-V(s)\right]P(0) & =\sum_{n=1}^{\infty}\frac{\Omega_{n}}{T^{n}}\\
\Omega_{n} & =\left.\mathcal{E}(s,\sigma)X_{n}(\sigma)W'(\sigma)\right|_{0}^{s}+\nonumber \\
 & -\int_{0}^{s}d\sigma\,\mathcal{E}(s,\sigma)X_{n}(\sigma)W''(\sigma)\nonumber \\
 & -\int_{0}^{s}d\sigma\,\mathcal{E}(s,\sigma)[PX_{n}'QW'](\sigma)\ .
\end{align}
In order to repeat the arguments for the case of level crossings, we
now have to assess the scaling of $PX_{n}'Q$ as one eigenvalue goes
to zero. We have separately shown that $X_{n}\sim1/\delta^{\beta_{n}}$
with $\beta_{n}=n\alpha+n-1$. Now, $X_{n}=S(X_{n-1}')$, so 
\begin{align}
PX_{n}'Q & =P(S(X_{n-1}'))'Q\\
 & =PS'X_{n-1}'Q\\
 & =PS'QX_{n-1}'Q\ ,
\end{align}
where the first equation holds because $PS=0$ and the second because
$PS'P=0$. Now, as we know, $PS'Q=-P'S$, introduces only a power $1/\delta^{\alpha}$.
On the other hand we know that $X_{n-1}'\sim1/\delta^{\beta_{n-1}+1}$.
Hence 
\begin{equation}
PX_{n}'Q\sim\frac{1}{\delta^{\beta_{n-1}+1+\alpha}}=\frac{1}{\delta^{\beta_{n}}}\ .
\end{equation}
We see hence that this potentially more dangerous term has the same
scaling as $X_{n}$. In other words, all our earlier conclusions also hold for the case with a degenerate kernel.

\section{Proof of an identity for $S'$} 
\label{app:identity}
Since $S=SQ$ [Eq.~\eqref{eq:S-props2}] we have
$S' = S' Q + S Q'$.
Note that
\beq
S Q' = - S P' = -S Q P' = -S^2 \mathcal{L} P' = S^2 \mathcal{L}' P \ ,
\eeq
where we used that $ S \mathcal{L} = Q$, and $\mathcal{L} P' = -\mathcal{L}' P$.  Differentiating $S\mathcal{L}=Q$
yields $S'\mathcal{L}+S\mathcal{L}'=-P'$. Multiplying from the
right by $S$ we obtain 
\beq
S'Q=-P'S-S\mathcal{L}'S\ .
\label{eq:S'Q1}
\eeq
Therefore we have:
\begin{eqnarray}
S' Q & =&  - P' Q S - S \mathcal{L}' S \nonumber \\
& =& -P' \mathcal{L} S^2 - S  \mathcal{L}' S = P \mathcal{L}' S^2 - S \mathcal{L}' S  \label{eq:dSQ}\ .
\end{eqnarray}
Combining the last three equations, we have:
\beq
S' = S^2 \mathcal{L}' P +  P \mathcal{L}' S^2 - S \mathcal{L}' S \ . \label{eq:Sprime}
\eeq

\section{Gap dependence for unitary families}
\label{sec:gap_unitary}

Assume here that $\mathcal{L}(s)$ is a unitary family, i.e., $\mathcal{L}(s)=e^{s\mathcal{K}}\mathcal{L}(0)e^{-s\mathcal{K}}$ with $\mathcal{K}$ an anti-hermitian superoperator. 
In this case one has (dropping the $s$ dependence) $\mathcal{L}'=\mathcal{K}\mathcal{L}-\mathcal{L}\mathcal{K}$
and so, using first order perturbation theory (\cite{kato_perturbation_1995}, page 77):
\begin{align}
P' &= - P\mathcal{L}' S - S \mathcal{L}' P \\
&=-P\mathcal{K}Q+Q\mathcal{K}P \\
  & = \mathcal{K}P - P \mathcal{K} \ .
\end{align}
%
Differentiating the above equation we get that both $P'$ and $P''$
are sums of products of $P$ and $\mathcal{K}$ and so are everywhere
bounded under the assumption that $P$ and $\mathcal{K}$ are bounded.
From Eq.~\eqref{eq:Sprime} and repeatedly using $S\mathcal{L} = \mathcal{L} S=Q$, $\mathcal{L}P=P\mathcal{L}=0$, and $S=SQ=SQ$ [from Eq.~\eqref{eq:S-props2}] we obtain
\begin{align}
S' &= S^2 (\mathcal{K}\mathcal{L}-\mathcal{L}\mathcal{K}) P +  P (\mathcal{K}\mathcal{L}-\mathcal{L}\mathcal{K}) S^2 - S (\mathcal{K}\mathcal{L}-\mathcal{L}\mathcal{K}) S \notag \\
& = -SQ\mathcal{K}P+P\mathcal{K}QS-S\mathcal{K}Q+Q\mathcal{K}S \notag \\
& = -S\mathcal{K}P+P\mathcal{K}S-S\mathcal{K}Q+Q\mathcal{K}S \notag \\
& = -S\mathcal{K} + \mathcal{K}S \ .
\end{align}
Hence in this case $S'$ carries only one inverse Liouvillian gap
(hidden in $S$). Combining these results we see that the constant
$C$ appearing in Eq.~(\ref{eq:constC}) is in this case at most $O(\Delta_{\mathrm{min}}^{-1})$.

\section{Bound on $\| P'\|$  for the KMS case}
\label{sec:bound_dP}

In the KMS case with a unique steady state we have already seen [Appendix~\ref{app:positivity}] that $P(s)x=\rho_{G}(s)\tr(x)$
with $\rho_{G}=e^{-\beta H}/Z$ and the partition function $Z=\tr e^{-\beta H}$. Then obviously
$P'(s)x=\rho_{G}'(s)\tr(x)$ and $\| P'\| =\| \rho'\| _{1}$.
We first recall the Bogoliubov-Duhammel scalar product \cite{Petz:1993fk,auerbach_interacting_1994}:
\begin{equation}
\left(A,B\right)=\frac{1}{Z}\int_{0}^{1}dx\,\tr\left[e^{-\beta(1-x)H}A^{\dagger}e^{-\beta xH}B\right],
\end{equation}
and define $\langle A\rangle_{G}:=\tr\left[Ae^{-\beta H}\right]/Z$.
Moreover one has the inequality  \cite{auerbach_interacting_1994}
\begin{equation}
\left(A,A\right)\le\frac{1}{2}\langle A^{\dagger}A+AA^{\dagger}\rangle \ .\label{eq:mermin_ineq}
\end{equation}
Differentiating the Gibbs state we obtain
\begin{equation}
\rho_{G}'=-\beta\frac{1}{Z}\int_{0}^{1}dx\, e^{-\beta(1-x)H}H'e^{-\beta xH}+\beta\rho_{G}\langle H'\rangle\ ,
\end{equation}
where the second term arises from the differentiation of the partition function. Now let $Y$ denote the first term of the above equation. We wish to bound its trace norm. We use \cite{Bhatia:book}
\begin{equation}
\| Y\| _{1}=\sup_{U}\left|\tr\left(YU\right)\right|\ ,
\end{equation}
where the supremum is taken over the set of unitary matrices $U$.
Note that $\tr\left(YU\right)=-\beta \left(H',U\right)$, since
$H'$ is Hermitian. Using the Schwartz inequality we have $\left|\tr\left(YU\right)\right|\le \beta \sqrt{\left(H',H'\right)\left(U,U\right)}$.
We now use Eq.~\eqref{eq:mermin_ineq} to obtain 
\begin{equation}
\| Y\| _{1}\le\beta\sqrt{\langle(H')^{2}\rangle_{G}}\ .
\end{equation}
On the other $\| \rho_{G}\| =1$. Combining the bounds
we get, all in all
\begin{equation}
\| \rho'\| _{1}\le\beta\left(\sqrt{\langle(H')^{2}\rangle_{G}}+\left|\langle H'\rangle_{G}\right|\right)\ .
\end{equation}
Since in general $\left\langle X^{2}\right\rangle >\left\langle X\right\rangle ^{2}$,
we can write an even more compact bound :
\begin{equation}
\| P'\| =\| \rho'\| _{1}\le2\beta\sqrt{\langle(H')^{2}\rangle_{G}}\ , \label{eq:dP_bound}
\end{equation}
as reported in the main text. 


\section{Extension to the case of level crossing}
\label{sec:level_crossing}

The results obtained so far assume that there is a finite gap from the 
zero eigenvalue to the rest of the spectrum. However, this is not necessary and all of the above can be generalized
to include the case in which there are level crossings along the adiabatic
path $[0,s]$. Suppose that there are $N$ level crossings at positions
$s_{i}^{*}$, $i=1,\ldots,N$. We isolate each level crossing by a
short segment of length $\delta_{i}$, so that no two segments overlap. Then define two sets $A,\, B$
such that $[0,s]=A\cup B$, with $A=\cup_{i}[s_{i}^{\ast}+\delta_{i}/2,s_{i+1}^{\ast}-\delta_{i+1}/2]$,
$i=0,\ldots,N$, with $s_{0}^{\ast}=0$, $s_{N+1}^{\ast}=s$, $\delta_{0}=\delta_{N+1}=0$,
and $B=\cup_{i}[s_{i}^{*}-\delta_{i}/2,s_{i}^{*}+\delta_{i}/2],$
$i=1,\ldots,N$. By construction, $A$ avoids the singularities while $B$ comprises short segments around each level crossing. We start with Eq.~\eqref{eq:start}, which we separate into integrals over $A$ and $B$:
\begin{multline}
\mathcal{E}^{-1}(s)V(s)P(0)-P(0)  =\int_{A\cup B}\mathcal{E}^{-1}W'd\sigma\\
  =-\frac{1}{T}\int_{A}(\mathcal{E}^{-1})'SW'd\sigma+\int_{B}\mathcal{E}^{-1}W'd\sigma\ ,
\end{multline}
where in the integral over $A$ we used Eq.~\eqref{eq:D3c}.
From the above equation, using the classic argument in Ref.~\cite{kato_adiabatic_1950}
one can than show that 
\begin{equation}
\lim_{T\to\infty}\left[\mathcal{E}(s)-V(s)\right]P(0)=0.
\end{equation}
However, we are interested in assessing the way in which the limit
is approached. In order to obtain a better estimate of the asymptotic behavior
it is necessary to repeat the integration by parts trick at all
orders. Since in region $A$ there are no exceptional points we can
immediately repeat all the steps from Appendix~\ref{sec:arb_order}. Assuming convergence of the 
series we then obtain
\begin{align}
&\left[\mathcal{E}(s)-V(s)\right]P(0)  =\sum_{n=1}^{\infty}\frac{\Omega_{n}^{A}}{T^{n}}-\int_{B}d\sigma\,\mathcal{E}(s,\sigma)W'(\sigma)\label{eq:series_ext-1}\\
&\Omega_{n}^{A}  =\sum_{i=0}^{N}\mathcal{E}(s,\sigma)X_{n}(\sigma)W'(\sigma)\Big|_{s_{i}^{\ast}+\delta_{i}/2}^{s_{i+1}^{\ast}-\delta_{i+1}/2} \nonumber \\
&\qquad \qquad -\int_{A}\!d\sigma\,\mathcal{E}(s,\sigma)X_{n}(\sigma)W''(\sigma) \ .\label{eq:seres_ext2-1}
\end{align}
Each term in the above equations is bounded for finite $\delta_{i}$,
however, when $\delta_{i}\to0$ some terms display singular behavior.
We are interested in estimating the size of the most singular terms.
Recall that we are assuming a Davies generator. In this case 
$W'=P'W$ [Eq.~\eqref{eq:W'}] is bounded (because $P'$ is bounded \cite{QAT-comment-smoothL} and $\| W\| =1$ because $W$ is a CPTP map),
and similarly $W''=[P''+(P')^{2}]W$ is bounded because $P''$ is bounded \cite{QAT-comment-smoothL}. Hence,
when $\delta_{i}\to0$, the only singularities in Eqs.~\eqref{eq:series_ext-1}-\eqref{eq:seres_ext2-1}
arise from $X_{n}(\sigma)$. Now assume that at each singular point $s_{i}^{\ast}$,
there is an eigenvalue  going to zero as $ v_{i}(s-s_{i}^{\ast})^{\alpha_{i}}$
(where the coefficients $v_{i}$ can be complex). This algebraic vanishing
of the gap is the only possibility in the finite-dimensional case
on which we focus. Moreover, for normal operators and analytic dependence
on $s$, the exponents $\alpha_{i}$ must be positive integers (see Ref.~\cite{kato_perturbation_1995}), but
we won't be needing this. Consider then the finite-dimensional
case. The normality of $\mathcal{L}$ guarantees that there are no nilpotent terms in the spectral resolution of $\mathcal{L}$ and so  the reduced
resolvent can be written as 
\begin{equation}
S(s)=\sum_{j>0}\frac{P_{j}(s)}{\lambda_{j}(s)}\ ,
\label{eq:S-sum}
\end{equation}
 where $\lambda_{j}(s)$ [$P_{j}(s)$] are the eigenvalues [spectral
projections] of $\mathcal{L}(s)$, with $P_{0}(s)=P(s)$. Hence
\begin{equation}
S'=-\sum_{j>0}\frac{P_{j}}{(\lambda_{j})^{2}}\lambda_{j}'+\sum_{j>0}\frac{P_{j}'}{\lambda_{j}}\ .
\label{eq:dS}
\end{equation}
Now, it is natural to assume that $P'_{j}$ are piecewise differentiable
also for $j>0$. For normal operators and analytic dependence on $s$
this follows from a Theorem of Kato (see Ref.~\cite{kato_perturbation_1995} Theorem
1.10, page 71). In this case the largest divergence of $S'$, when
$s-s_{i}^{\ast}=\delta_{i}\to 0$, comes from the vanishing of the
gap in the first term of Eq.~(\ref{eq:dS}) and is of the form
\begin{equation}
S'(s)\sim\frac{1}{\delta_{i}^{\alpha_{i}+1}}\ .
\end{equation}
Note that the prefactors may be different depending on whether $\delta_{i}\to0^{\pm}$.
This reasoning can be extended to $X_{n}(s)$. Recall that $X_{n}  =S X_{n-1}'$, with $X_{1}=S$, and note that $X_{n}$
contains $n$ powers of $S$ and a total of $n-1$ derivatives with
respect to $s$. So we conclude that $X_{n}(s)\sim1/\delta_{i}^{n\alpha_{i}+n-1}$
as $\delta_{i}\to0.$ More precisely we have
\begin{equation}
\lim_{\delta_{i}\to0^{\pm}}\delta_{i}^{\beta_{n}^{i}}X_{n}(s_{i}^{\ast}+\delta_{i})=Y_{n}^{\pm}\ ,
\end{equation}
where the $Y_{n}^{\pm}$ exist and where 
\beq
\beta_{n}^{i}=n\alpha_{i}+n-1\ .
\eeq 
Thanks to the above there exist positive
constants $A_{n}^{i}$ such that, for sufficiently small $\delta_{i}$
\begin{equation}
\| \Omega_{n}^{A}\| \le\sum_{i=1}^{N}\frac{A_{n}^{i}}{\delta_{i}^{\beta_{n}^{i}}}\ .
\end{equation}
The $A_{n}^{i}$ can be made independent of $\delta_{i}$, for sufficiently
small $\delta_{i}$. In fact they can be taken to be the max of the
norm of the most diverging terms plus a small constant. On the other
hand, to bound the second term of Eq.~(\ref{eq:series_ext-1}) we
just need to notice that $W'$ is piecewise continuous
\begin{multline}
\| -\int_{B}\!d\sigma\mathcal{E}(s,\sigma)W'(\sigma)\|  \\=\| -\sum_{i=1}^{N}\int_{s_{i-\delta_{i}/2}^{*}}^{s_{i+\delta_{i}/2}^{*}}\!d\sigma\mathcal{E}(s,\sigma)W'(\sigma)\| 
  \le\sum_{i=1}^{N}\delta_{i}B_{i} \ ,
\end{multline}
with positive $B_{i}$, because the integrand is piecewise continuous
($B_{i}\sim(1/2)(\| P'(s_{i}^{*}+0^{+})\| +\| P'(s_{i}^{*}-0^{+})\| )$.
All in all we obtained
\beq
\| \left[\mathcal{E}(s)-V(s)\right]P(0)\|   \le\sum_{i=1}^{N}f_{i}(T,\delta_{i})\ ,
\label{eq:le_fi}
\eeq
where
\beq
f_{i}(T,\delta_{i}) =\sum_{n=1}^{\infty}\frac{A_{n}^{i}}{T^{n}\delta_{i}^{\beta_{n}^{i}}}+B_{i}\delta_{i}\ .
\label{eq:f_start-1}
\eeq
Now for each fixed $T$, we are free to minimize each $f_{i}$ over
$\delta_{i}$, to which we turn next. 

Differentiating, we obtain
\begin{equation}
\partial_{\delta_{i}}f_{i}=-\sum_{n=1}^{\infty}\frac{\beta_{n}^{i}A_{n}^{i}}{T^n\delta_{i}^{\beta_{n}^{i}+1}}+B_{i}=0\ ,
\label{eq:df}
\end{equation}
where $\beta_{n}^{i}+1=n(\alpha_{i}+1)$. Recall that all the $A_{n}^{i}\ge0$
and $B^{i}>0$, so the only way Eq.~\eqref{eq:df} can have a solution
is that, asymptotically as $T\to\infty$, some of the terms in the
series in Eq.~\eqref{eq:df} scale to constant, while some others
may scale to zero. Plugging in the Ansatz $\delta_{i}=(c_{i}T)^{-\eta_{i}}$
one can check that $\eta_{i}=1/(\alpha_{i}+1)$ is the only possible
solution. In fact all the terms scale to a constant. The constants $c_{i}$
must satisfy
\begin{equation}
\sum_{n=1}^{\infty}\beta_{n}^{i}A_{n}^{i}c_{i}^{n}=B_{i} \ .
\end{equation}
The extremum is indeed a minimum as it is easy to check that the second
derivative is always positive for positive $T,\delta$. Substituting the
solution back into Eq.~\eqref{eq:f_start-1} we obtain
\begin{align}
f_{i}(T,\delta_{\mathrm{min}}) & =\frac{D_{i}}{(c_{i}T)^{\eta_{i}}}\\
D_{i} & =\left[\sum_{n=1}^{\infty}A_{n}^{i}(c_i)^{n}+B_{i}\right].
\end{align}

Returning to Eq.~\eqref{eq:le_fi} we finally obtain 
\begin{equation}
\| \left[\mathcal{E}(s)-V(s)\right]P(0)\| \le\sum_{i=1}^{N}\frac{D_{i}}{(c_{i}T)^{\eta_{i}}} \ ,
\end{equation}
as stated in Eq.~(5) 
 of the main text.

\section{Derivation of the eigenvalues of the Lindbladian of Example 2}
\label{app:ex2-deriv}

Denote the instantaneous Hamiltonian gap by $\delta(s)$ and the instantaneous ground state and first excited state by $\ket{\epsilon_0(s)}$ and $\ket{\epsilon_1(s)}$ respectively.  The action of the Lindbladian on $\rho$ can be written in this basis as (dropping the $s$ dependence for clarity):
\bes
\begin{align}
\bra{\epsilon_0}\mathcal{L} \rho \ket{\epsilon_0} & =  \gamma(\delta) \left[ |A_{01}|^2 \rho_{11} -  e^{-\beta \delta} |A_{10}|^2 \rho_{00} \right] \\
\bra{\epsilon_0}\mathcal{L} \rho \ket{\epsilon_1} & =   \left( i \mu - \Gamma  \right) \rho_{01} \\
\bra{\epsilon_1}\mathcal{L} \rho \ket{\epsilon_0} & =   \left( -i \mu - \Gamma  \right) \rho_{10} \\
\bra{\epsilon_1}\mathcal{L} \rho \ket{\epsilon_1} & =  \gamma(\delta) \left[ e^{-\beta \delta}  |A_{10}|^2 \rho_{00}  - |A_{01}|^2 \rho_{11} \right]\ ,
\end{align}
\ees
where $\rho_{ab} \equiv \bra{\epsilon_a} \rho \ket{\epsilon_b}$, $\mu = \delta -S(0) (A_{00}^2 -A_{11}^2 ) + [S(\delta) -S(-\delta)] |A_{01}|^2$ and $2\Gamma=\gamma(0)(\left|A_{00}\right|^{2}+\left|A_{11}\right|^{2})+\gamma(\delta)\left|A_{01}\right|^{2}(1+e^{-\beta\delta})$.
This form immediately allows us to read off two of the eigenvalues of $\mathcal{L}$ as $\pm i \mu - \Gamma$.  The remaining two are the eigenvalues of the following $2 \times 2$ matrix:
\beq
\gamma(\delta)|A_{10}|^2 \left(
\begin{array}{cc}
 - e^{-\beta \delta}   & 1 \\
 e^{-\beta \delta}  & -1
\end{array}
\right)\ ,
\eeq
whose eigenvalues are $0$ and $ -|A_{01}|^2 \gamma(\delta)  \left( 1 + e^{-\beta \delta} \right) $.
%


\begin{thebibliography}{54}
\bibitem [{QAT()}]{QAT-comment-smoothL}%
  \BibitemOpen
  \href@noop {} {}\bibinfo {note} {{Throughout we require the technical
  assumption that $\mathcal{L}(s)$ depends sufficiently smoothly on $s$, which
  together with the gap condition implies that $P(s)$ has sufficiently many
  bounded derivatives (e.g., $\mathcal{L}(s)$ and $P(s)$ are analytic for
  $s\in[0,1]$). The reduced resolvent is then bounded and smooth.}}\BibitemShut
  {Stop}%
\end{thebibliography}
%

\end{document}